\documentclass[english]{article}
\usepackage{graphicx}
\pdfoutput=1
\usepackage{changepage}

\usepackage[utf8x]{inputenc}

\usepackage{textcomp,marvosym}

\usepackage{cite}

\usepackage{nameref,hyperref}

\usepackage[right]{lineno}

\usepackage{microtype}
\DisableLigatures[f]{encoding = *, family = * }

\usepackage[table]{xcolor}

\usepackage{array}

\newcolumntype{+}{!{\vrule width 2pt}}

\newlength\savedwidth

\newcommand\thickhline{\noalign{\global\savedwidth\arrayrulewidth\global\arrayrulewidth 2pt}%
\hline
\noalign{\global\arrayrulewidth\savedwidth}}

\usepackage[T1]{fontenc}

\usepackage{amsmath,amssymb}
\usepackage{graphicx}
\usepackage{cite}



\begin{document}

\title{Long-range memory test by the burst and inter-burst duration distribution}

\author{Vygintas Gontis}

\date{Institute of Theoretical Physics and Astronomy, Vilnius University, Vilnius, Lithuania, E-mail: vygintas@gontis.eu}
\maketitle
\vspace{10pt}

\begin{abstract}
It is empirically established that order flow in the financial markets is positively auto-correlated and can serve as an example of a social system with long-range memory. Nevertheless, widely used long-range memory estimators give varying values of the Hurst exponent. We propose the burst and inter-burst duration statistical analysis as one more test of long-range memory and implement it with the limit order book data comparing it with other widely used estimators. This method gives a more reliable evaluation of the Hurst exponent independent of the stock in consideration or time definition used. Results strengthen the expectation that burst and inter-burst duration analysis can serve as a better method to investigate the property of long-range memory.  
\end{abstract}

%
\vspace{2pc}

%
%

\section{Introduction \label{sec:introduction}}
The long-range memory in natural and social systems ranges from the level of rivers to the financial markets. The vast amount of empirical data and observed power-law statistical properties of volatility and trading activity in the financial markets are still among the most mysterious features attracting the permanent attention of researchers \cite{Baillie1996JE,Engle2001QF,Plerou2001QF,Gabaix2003Nature,Ding2003Springer}. From our point of view, the definition of long-range memory based on the self-similarity and power-law statistical properties are ambiguous as Markov processes can exhibit all these properties, including slowly decaying autocorrelation \cite{Gontis2004PhysA,McCauley2006PhysA,McCauley2007PhysA,Micciche2009PRE,Micciche2013FNL,Ruseckas2011PRE}. Econometricians tend to conclude that the statistical analysis, in general, cannot be expected to provide a definite answer concerning the presence or absence of long-range memory in asset price return \cite{Lo1991Econometrica,Willinger1999FinStoch,Mikosch2003}. Nevertheless,  it is widely accepted that the volatility of prices exhibits long-range memory properties. Thus the alternative models, such as FIGARCH, FIEGARCH, LM-ARCH and ARFIMA, including fractional Brownian noise (fBn), have been proposed for the volatility in the financial markets \cite{Ding1993JEmpFin,Baillie1996JE,Bollerslev1996Econometrics,Giraitis2009,Conrad2010,Arouri2012,Tayefi2012}.  

Earlier we have proposed an agent-based model, macroscopic dynamics of which can be reduced to  a set of stochastic differential equations (SDEs) able to reproduce empirical probability density function (PDF) and power spectral density (PSD) of absolute return \cite{Kononovicius2013EPL,Gontis2014PlosOne} as well as scaling behavior of volatility return intervals \cite{Gontis2016PhysA}. Later, investigating empirical PDF of burst and inter-burst duration compared with the model properties we have explained the so-called long-range memory in the financial markets by ordinary non-linear SDEs representing multifractal stochastic processes with non-stationary increments \cite{Gontis2017PhysA,Gontis2018PhysA}. The proposed description is an alternative to the modeling incorporating fractional Brownian motion (fBm) and might be applicable in the modeling of other social systems, where models of opinion or populations dynamics lead to the macroscopic description by the non-linear SDEs \cite{Gontis2017Entropy}. 

From our perspective, there is a fundamental problem empirically establishing which of the possible alternatives, fBm or stochastic processes with non-stationary increments, is most well-suited to describe natural and social systems exhibiting power-law statistical properties and self-similarity. Our first idea was to employ the dependence of first passage time PDF on Hurst exponent $H$ for the fBm \cite{Ding1995PhysRevE,Metzler2014Springer,Gontis2017PhysA,Gontis2018PhysA} in the empirical analysis of long-range memory properties for the volatility in the financial markets. 

As was shown in the empirical analysis of order flow in the financial markets, there is strong evidence of the persistence of the order signs \cite{Lillo2004SNDE,Bouchaud2004QF,Toth2015JEDC}. Authors employed the main statistical methods to evaluate the Hurst exponent of the order flow time series and have found that, in most cases $H\simeq 0.7$. Here we investigate burst and inter-burst duration statistical properties of order dis-balance time series seeking to confirm or reject the long-range memory in the order flow.

The rest of this paper is organized as follows. First, we describe methods used to evaluate the property of long-range memory. Second, we describe used data sources of limit order books (LOB) and define  order dis-balance time series. Further, we apply our methods to the data and demonstrate the advantages of burst and inter-burst duration analysis.  In the concluding part, we discuss the results and summarize findings. 

\section{Methods \label{sec:methods}}
The most widespread definition of long-range memory is based on the power-law autocovariance 
\begin{equation}
\gamma(\tau)\sim \tau^{-\alpha} L(\tau)
\label{eq:memory}
\end{equation}
with divergent integral in the limit $\tau\rightarrow\infty$, where $0<\alpha<1$ and $L(x)$ is a slowly varying function at infinity, $\lim_{x\rightarrow \infty}L(tx)/L(x)=1$. This power-law behavior of autocovariance is related to the self-similarity and other power-law statistical properties. Thus, the Hurst exponent $H$, parameter of self-similarity, is also essential in long-range memory theory, where $H=1-\alpha/2$. For the short-memory Markov processes $H=1/2$ and the positively correlated long-memory process $0.5<H<1$. This definition of long-range memory is ambiguous, when one deals with the real-time series finite in length, as Markov processes in some regions of variables can generate power-law statistical properties, including power-law auto-correlation as well. Thus we distinguish two  different cases of long-range memory: a) true
long-range memory process, one with correlated increments, such as the fractional Brownian
motion (fBm) \cite{Mandelbrot1968SIAMR,McCauley2006PhysA,McCauley2007PhysA}, and b) the spurious long-range memory process, such as Markov processes with non-stationary uncorrelated increments \cite{Gontis2004PhysA,Kaulakys2005PhysRevE,Micciche2009PRE,Ruseckas2011PRE,Micciche2013FNL}. The cases of spurious long memory were considered by Lanouar as well \cite{Lanouar2011IJBSS}.   

The Hurst exponent is a universal parameter of self-similarity. It defines the relation of the standard deviation to the time $t$, $t^H L(t)$, where $H=1/2$ only in the cases when stochastic increments are not correlated. Yet another  equivalent definition of the long-range memory is used in terms of the spectral density for low frequencies
\begin{equation}
S(f)=f^{1-2H} LL(f),
\label{eq:spectral-density}
\end{equation}
where $f$ is the frequency, and $LL(f)$ is a slowly varying function in the limit $f\rightarrow 0$. Notice that this last condition has the same constrain related to the requirement of time series infinite in time. 

There are many different methods to evaluate $H$ from empirical time series. Authors seek to combine different methods to increase the reliability of results. In this paper we will use three well known Hurst exponent estimators described bellow in comparison with our estimation from the exponent of burst and inter-burst duration PDF, see for example \cite{Gontis2017PhysA,Gontis2017Entropy,Gontis2018PhysA}.

The first method. Estimation of the spectral density $S(f)$ using the periodogram method 
\begin{equation}
S(f)=\frac{1}{2 \pi t_n}\vert \sum_{j=1}^{j=n} X_j e^{i t_j f} \vert^2,
\label{eq:periodogram}
\end{equation}
where $t_j$ is the time of event -- occurrence of value change in the time series $X_j$ and $t_n$ is the length of time series. We evaluate the exponent $1-2H$ from the power-law \eqref{eq:spectral-density} expected for the low values of the frequency.

The second method is the rescaled range (R/S) method \cite{Beran1994Chapman}. R/S is a widely used and described method, in summary, one can divide a time series $X_1,X_2,...,X_N$ with equal $N-1$ time steps into $m$ subseries each of length $n,(m \cdot n=N)$, mean $e_j$ and standard deviation $S_j$. Then define cumulative deviations from the mean $x_{k,j}$, the range $R_j$, and the value of rescaled range $(R/S)_n$
\begin{align}
x_{k,j} &=\sum_{i=1}^{i=k}(X_{i,j}-e_j),\nonumber\\
R_j &=max_k(x_{k,j})-min_k(x_{k,j}),\\
(R/S)_n &=\frac{\sum_{j=1}^m R_j/S_j}{m}.\nonumber
\label{eq:rescaled-range}
\end{align}
We repeat the procedure of $(R/S)_n$ calculation for the consecutive $n$, dividing the series of data into disjoint intervals and finding the mean value $(R/S)_n$ for each value of $n$. Finally, we calculate the Hurst exponent $H$ as the slope in the log-log graph of $\lg (R/S)_n$ versus $\lg n$.

The third method is the multifractal detrended fluctuation analysis (MF-DFA) \cite{Kantelhardt2002PhysA} as a generalized version of detrended fluctuation analysis (DFA) \cite{Peng1994PRE}. First, we integrate the initial time series of length $N$, denote it $X_t$, and divide into $m$ boxes of equal length $n$. In each box $j$, a least-squares line is fit to the data giving us $Y_n(k,j)$ coordinates of the straight line segments. Then calculate variances of deviation from the each trend line $F^2(n,j)=\frac{1}{n} \sum_{k=1}^m (X_n(k,j)-Y_n(k,j))^2$. Average overall segments to obtain the q-th order
fluctuation function
\begin{equation}
F_q(n)=\{\frac{1}{m} \sum_{j=1}^m \left[F^2(n,j)\right]^{\frac{q}{2}}\}^{\frac{1}{q}}.
\label{eq:q-fluctuations}
\end{equation}
Finally we calculate the generalized Hurst exponent $H_q$ as the slope in the log-log graph of $\lg F_q(n)$ versus $\lg n$.

\subsection{Data \label{subsec:data}}
We use financial data as the best available time series of a social systems with expected true long-range memory property. High-frequency, easy to use limit order book data for all NASDAQ traded stocks is provided by Limit Order Book System LOBSTER \cite{Huang2011Lobster}. The limit order book (LOB) data that LOBSTER reconstructs originates from NASDAQ's Historical TotalView-ITCH files (http://nasdaqtrader.com). We have selected only five stocks for this research as LOBSTER provides AAPL, AMZN, GOOG, INTLC, MSFT as free
access sample files of data for these stocks. We included two distant periods of
records aiming to detect structural changes in the market, which might have
occurred. The first period covers all trading dates from 26 of June to 31 of August
in 2012 and second period covers days from 6 of January to 28 of February in
2020. When analysis concerns the data in 2020, we add the year to the stock
symbol, such as AAPL-2020. The first period covers 48 trading days and
the second 38 days.

LOBSTER generates two files: message.csv and orderbook.csv for each selected trading day and ticker (stock). The message.csv file contains the full list of events causing an update of LOB in the requested price range. We include orders up to the 10 levels of prices for this research. All events are time-stamped in seconds with a precision of at least milliseconds and up to the nanoseconds depending on the selected period. Both files provide exact information about the instantaneous state of LOB needed to define order dis-balance time series. Any event $j$ changing the LOB state has a time record $t_j$ and 10 values of buy volumes $v_k^{+}(t_j)$ as well as sell volumes $v_k^{-}(t_j)$, which are retrieved from LOB data at any moment. We define order dis-balance time series $X(t_j)$ as follows
\begin{equation}
X(t_j)=\frac{\sum_{k=1}^{10}(v_k^{+}(t_j)-v_k^{-}(t_j))}{\sum_{k=1}^{10}(v_k^{+}(t_j)+v_k^{-}(t_j))}.
\label{eq:order-disbalance}
\end{equation}
It follows from the definition that order dis-balance time series is bounded and fluctuates in the interval $-1 \leq X(t_j) \leq 1$. It is reasonable to consider that so defined order dis-balance is a measure of traders opinion about the price and this empirical time series is an example of opinion dynamics in the social system.

For the application of statistical methods, we  have to combine time series prepared from daily data into series up to the few months. Thus we drop small parts of daily series $X(t_j)$ at the very beginning of the day and at the end to ensure that daily series start and end with the value $X(t_j) \simeq 0$. Then we combine these successive daily intervals into continuing order dis-balance time series of the needed length.

\begin{figure}
\begin{centering}
\includegraphics[width=0.95\textwidth]{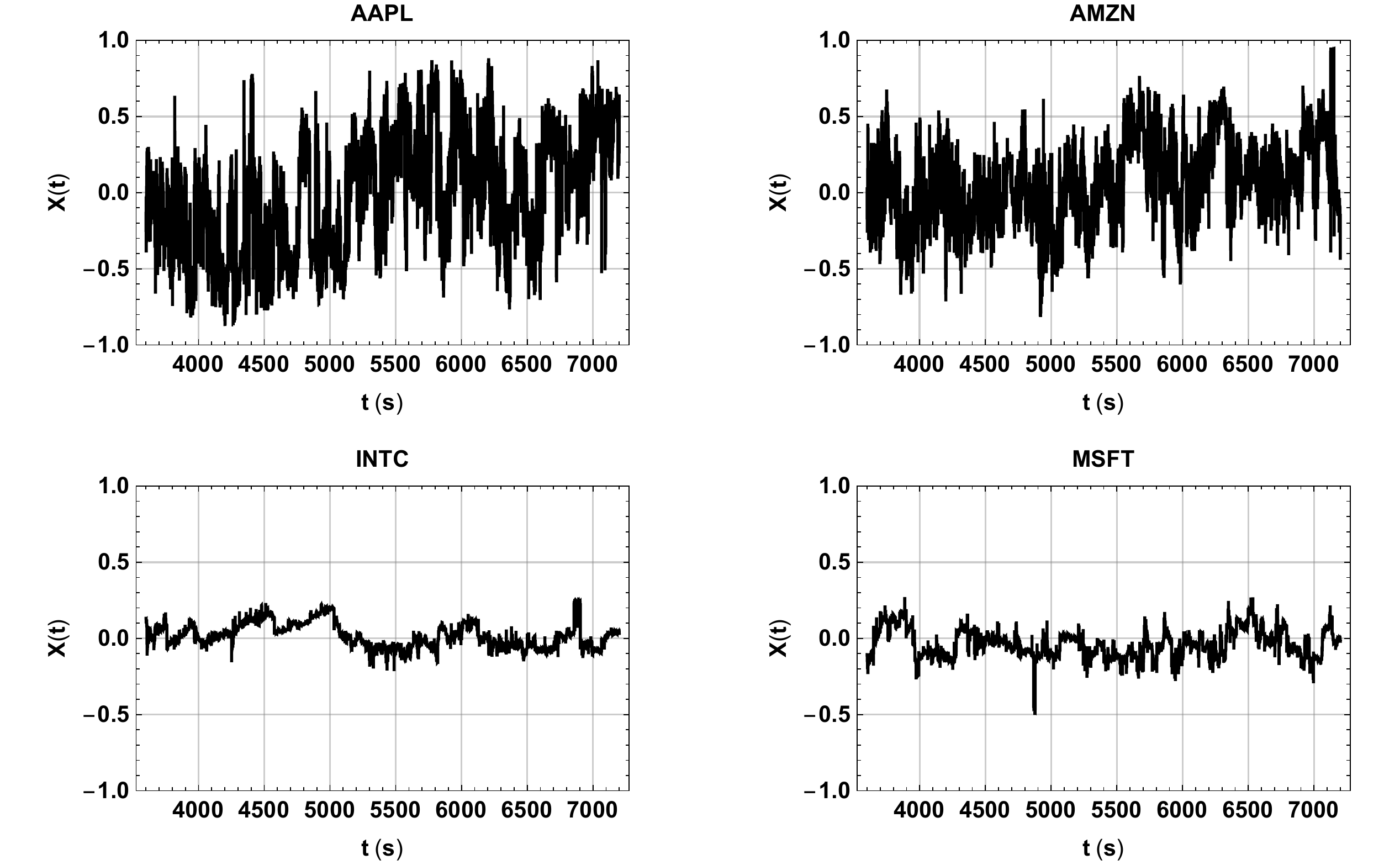}
\par\end{centering}
\caption{Examples of order dis-balance real-time series.
The excerpts of the second hour of order dis-balance real-time series for four stocks: AAPL, AMZN, INTC, MSFT traded on the June 21 of 2012 and reconstructed from the LOBSTER sample data. \label{fig1}}
\end{figure}

For this research, we use two types of time series: when the time is measured in seconds as retrieved from data and when the time is measured in ticks of events.  The event time changes by one tick when any change of order book state appears. In figure \ref{fig1} we demonstrate the one trading hour (the second in the trading day) excerpt of  order dis-balance real-time series. One can observe that the range of fluctuations is varying for different stocks. The intensity of order flow $\nu$ measured, for example, in the number of events per hour, is varying as well.  Values $\nu$ for the stocks AAPL, AMZN, GOOG, INTC, MSFT are \{60752, 49903, 25868, 87163, 106258\}.

\subsection{Burst and inter-burst duration \label{subsec:duration}}

\begin{figure}
\begin{centering}
\includegraphics[width=0.5\textwidth]{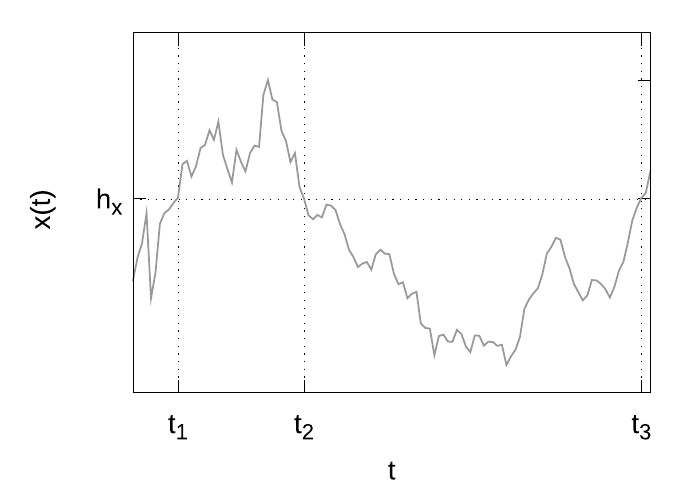}
\par\end{centering}
\caption{Definition of burst and inter-burst duration.
Excerpt from a generic time series $X(t)$. Three threshold, $h_x$, passage events, $t_i$, are shown. Thus burst duration can be defined as $t_2 − t_1$, and inter-burst duration can be defined as $t_3 − t_2$. \label{fig2}}
\end{figure}

The most innovative part of this contribution is related to the burst and inter-burst duration statistical analysis. The scientific motivation to work on the empirical analysis of burst duration statistics comes from the need for criteria to discriminate between true long-range memory processes such as fBm and spurious long-range memory as modeled by the non-linear SDEs  \cite{Gontis2017PhysA,Gontis2017Entropy,Gontis2018PhysA}. The definition of burst and inter-burst duration is given in figure \ref{fig2}. There are two
distinct threshold passage events — one describes a return to the threshold from
above, while the other describes a return to the threshold from below. Burst
duration measures the time series spent above the threshold (starts with a passage
from below and ends with a passage from above; see $t_2 − t_1$ in figure \ref{fig2}). While the
inter-burst duration measures the time series spent below the threshold (starts
with a passage from above and ends with a passage from below;
see $t_3 − t_2$ in figure \ref{fig2}). We considered the concept of burst and inter-burst duration analysis more extensively in  \cite{Gontis2017PhysA,Gontis2017Entropy}. In this contribution, we use the only notation $T$ for burst and inter-burst duration as differences are not relevant here.

The idea of how to discriminate is based on the PDF of burst and inter-burst duration $T$ defined for the fBm \cite{Ding1995PhysRevE,Metzler2014Springer,Gontis2017PhysA,Gontis2018PhysA} 
\begin{equation}
P(T)\sim T^{H-2}.
\label{eq:T-PDF}
\end{equation}
Power-law \eqref{eq:T-PDF} with exponent $\gamma=2-H$ is defined by the Hurst exponent $H$ and any deviation from the exponent $3/2$ or $H=1/2$ should indicate the presence of correlations in the noise increments and true long-range memory. Markov processes should always give us at least a part of a burst duration PDF as power-law with exponent $3/2$. 
In the previous work, we have shown that non-linear SDEs, being a  macroscopic description of many agent-based systems or birth-death processes, generate time series with power-law $3/2$ \cite{Gontis2012ACS,Gontis2014PlosOne,Gontis2016PhysA,Gontis2017Entropy,Gontis2018PhysA}.  
The non-linear SDEs are in the background of the stochastic models of trading activity and volatility in the financial markets and explain power-law statistical and long-range memory properties, including power-spectral density and auto-correlation \cite{Gontis2006JStatMech,Gontis2007PhysA,Kononovicius2012PhysA,Gontis2014PlosOne}. This type of spurious long-range memory is a very realistic alternative to the modeling incorporating fBm. The empirical  burst duration analysis (BDA) in the order dis-balance time series should give us an answer about whether the origin of long-range memory in the order flow is spurious or true. 

It is worth noting that the proposed burst duration analysis is
based on the assumption of a one-dimensional stochastic process with the power-law \eqref{eq:T-PDF}. The exponent of PDF, in this case, is in the interval  $1 \leq \gamma \leq 2$. The power-law we expect in the general case has to appear in the middle part of PDF when the exact form of PDF on both sides is usually unknown. Thus non-parametric methods such as \cite{Clauset2009SIAMR} are not applicable for this analysis. We have just to rely on the assumption of power-law. There are more complex cases in nature, see, for example,
empirical analyses of earthquake interoccurrence time distribution \cite{Corral2006Tectonophysics,Lennartz2008EPL}. The
Gamma distribution gives an excellent fit to the earthquake interoccurrence time
PDF, but the power-law exponent, in this case, is considerably lower than 1. Probably
the stochastic nature of inter-event time and magnitude makes this case less
applicable for the burst duration analysis we discuss in this contribution.

\section{Results \label{sec:reults}}
\begin{figure}
\begin{centering}
\includegraphics[width=0.95\textwidth]{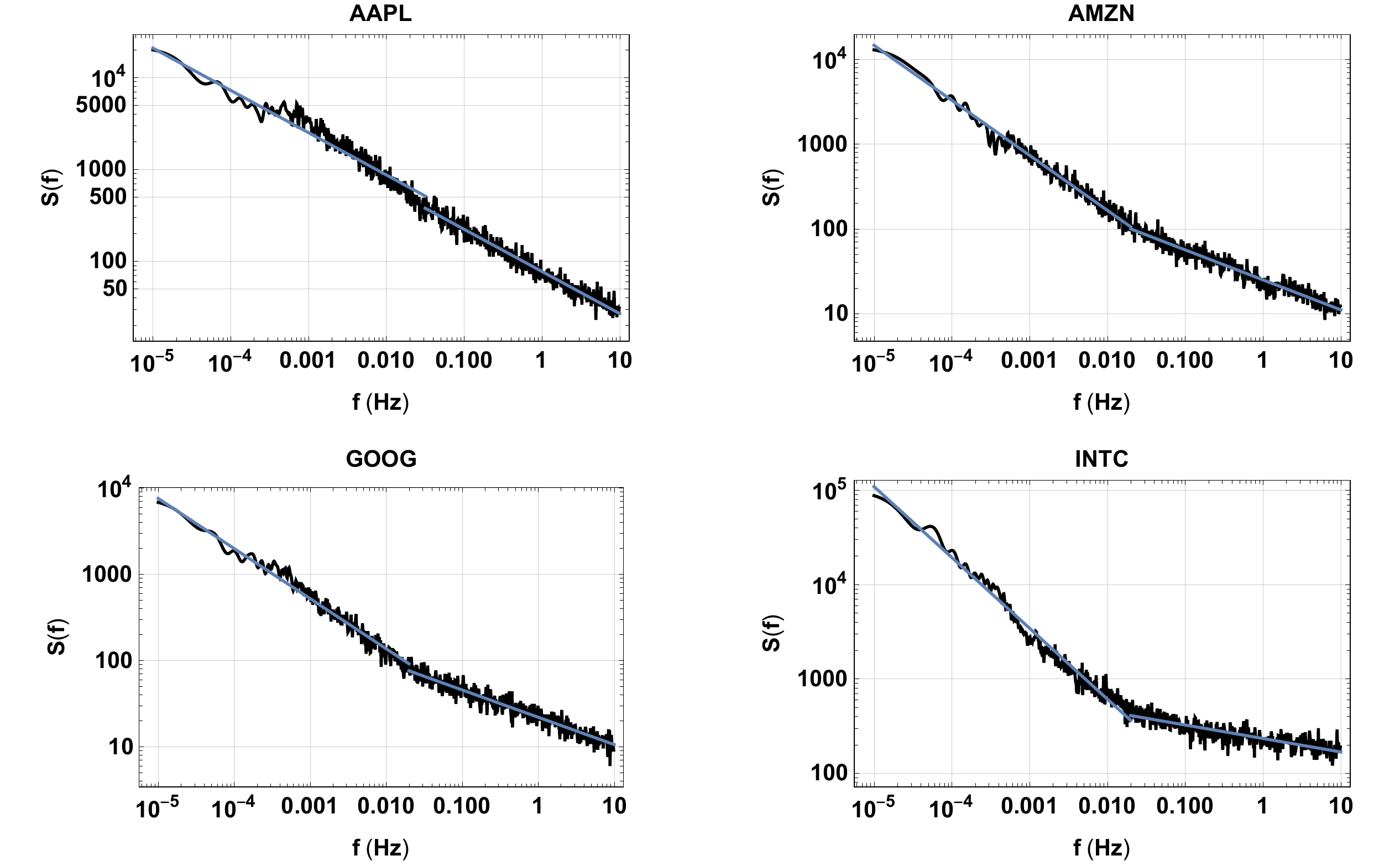}
\par\end{centering}
\caption{Power spectral density of order dis-balance real-time  series.
The spectra are calculated by periodogram \eqref{eq:periodogram} for the four stocks: AAPL, AMZN, GOOG, INTC and is averaged over 48 trading days. Straight lines fit the PSD given on the log-log scale. \label{fig3}}
\end{figure}

We consider the order dis-balance time series retrieved from LOBSTER data as described in the previous sub-section and compare the statistical properties of real-time and event time series. First of all, the complexity of these financial time series is revealed in the power spectral density (PSD) $S(f)$ calculated from the periodogram \eqref{eq:periodogram}. In figure \ref{fig3} we demonstrate averaged PSD of real-time series calculated for the four stocks: AAPL, AMZN, GOOG, INTC. The average of spectra here is calculated from the 48 daily series in the period from June 26 to August 31 of 2012. 

Note that the PSD is fractured and one can define at least two power-law $S(f)\sim 1/f^{\beta}$ exponents: $\beta_1$ for the lower frequencies and $\beta_2$ for the higher. From first glance, this property looks similar to the empirical PSD of the return volatility in the financial markets \cite{Gontis2014PlosOne,Gontis2016PhysA}. In table \ref{table1} we present values of exponents evaluated by the log-log linear fit of PSDs for the five stocks, see fitting  lines given in figure \ref{fig3}. Here we in parentheses give values calculated from the 38 daily series from January 6 to February 28 of 2020.

\begin{table}
\centering
\caption{
Power-law exponents of the power spectral density for the five stocks order dis-balance time series. Values in parenthesis are calculated from the time series of 2020.}
\begin{tabular}{|l+l|l|l|l|l|l|l|}
\hline
\multicolumn{1}{|l|}{\bf Exponents} & \multicolumn{1}{|l|}{\bf AAPL} & \multicolumn{1}{|l|}{\bf AMZN} & \multicolumn{1}{|l|}{\bf GOOG} & \multicolumn{1}{|l|}{\bf INTC} & \multicolumn{1}{|l|}{\bf MSFT}\\ \thickhline
$\beta_1$ & $0.46(0.54)$ & $0.65(0.69)$ & $0.58(0.69)$ & $0.75(0.72)$ & $0.60(0.62)$ \\ \hline
$\beta_2$ & $0.46(0.34)$ & $0.35(0.20)$ & $0.31(0.36)$ & $0.14(0.22)$ & $0.17(0.25)$ \\ \hline
$H$ & $0.73(0.77)$ & $0.83(0.85)$ & $0.79(0.85)$ & $0.88(0.86)$ & $0.80(0.81)$ \\ \hline
\end{tabular}
\label{table1}
\end{table}
The straightforward interpretation of PSD and its exponents by the relation \eqref{eq:spectral-density} becomes complicated when we have fractured PSD with different values of $\beta$ for the various stocks and frequencies. The fluctuations in order flow intensity are probably important for the higher frequencies and contribute to PSD's behavior. In table \ref{table1} we present the formal estimation of $H=(1+\beta_1)/2$, as PSD for lower frequencies has to define long-range memory properties. Though this estimation of $H$ gives us considerable deviation form other methods given below, the further analysis of inter-burst duration PDF will strengthen the choice of lower frequencies. We do not observe considerable fluctuations of PSD from one day to another for the same stock, and changes from one time period to another are not so considerable as fluctuations among various stocks. For the stocks considered, the peculiarities of PSD look like the characteristic feature of the asset.  
Considerable fluctuations in the slopes of PSDs for the various stocks and time scales implies the need to evaluate Hurst exponent using other methods. 

The second, R/S method to evaluate Hurst exponent gives more definite results than PSD. We chose the same time steps $\tau=200 \: s$ in the implementation of R/S method for the order dis-balance real-time series. In the case of the event time series we used $\tau_1=500 \: ticks$ for the AAPL, AMZN, GOOG stocks and $\tau_2=2000 \: ticks$ for the INTC and MSFT stocks. We provide  the list of evaluated $H$ values in table \ref{table2}, where the values in parenthesis are from the year 2020 series.
\begin{table}
\centering
\caption{
Hurst exponent $H$ evaluated by R/S and MDFA methods for the five stocks order dis-balance real and event time series. Values in parenthesis are calculated from the time series of 2020.}
\begin{tabular}{|l+l|l|l|l|l|l|l|}
\hline
\multicolumn{1}{|l|}{\bf Exponents} & \multicolumn{1}{|l|}{\bf AAPL} & \multicolumn{1}{|l|}{\bf AMZN} & \multicolumn{1}{|l|}{\bf GOOG} & \multicolumn{1}{|l|}{\bf INTC} & \multicolumn{1}{|l|}{\bf MSFT}\\ \thickhline
$H, 200 \: s$ & $0.64(0.62)$ & $0.74(0.72)$ & $0.69(0.69)$ & $0.78(0.71)$ & $0.75(0.73$ \\ \hline
$H, 500 $  & $0.59(0.72)$ & $0.66(0.75)$ & $0.61(0.78)$ &  &  \\ \hline
$H, 2000 $  & &  & & $0.78(0.80)$ & $0.72(0.79$\\ \thickhline
$H_2, 200 \: s$ & $0.59(0.73)$ & $0.57(0.68)$ & $0.55(0.63)$ & $0.64(0.61)$ & $0.52(0.62)$ \\ \hline
$H_2, 500 $  & $0.60(0.64)$ & $0.59(0.67)$ & $0.58(0.63)$ &  &  \\ \hline
$H_2, 2000 $  & &  & & $0.64(0.61)$ & $0.58(0.59)$\\ \hline
\end{tabular}
\label{table2}
\end{table}

Though the values of $H$ in table \ref{table2} are scattered, these numbers are compatible with the estimation by PSD, other order flow empirical researches \cite{Lillo2004SNDE,Bouchaud2004QF,Mike2008JEDC,Toth2015JEDC} and confirm the presence of long-range memory. Values of $H$ for the event time series look slightly lower than for the real-time, but in both cases, they are scattered around $H=0.7$ and confirm findings of other researches.

\begin{figure}
\begin{centering}
\includegraphics[width=0.95\textwidth]{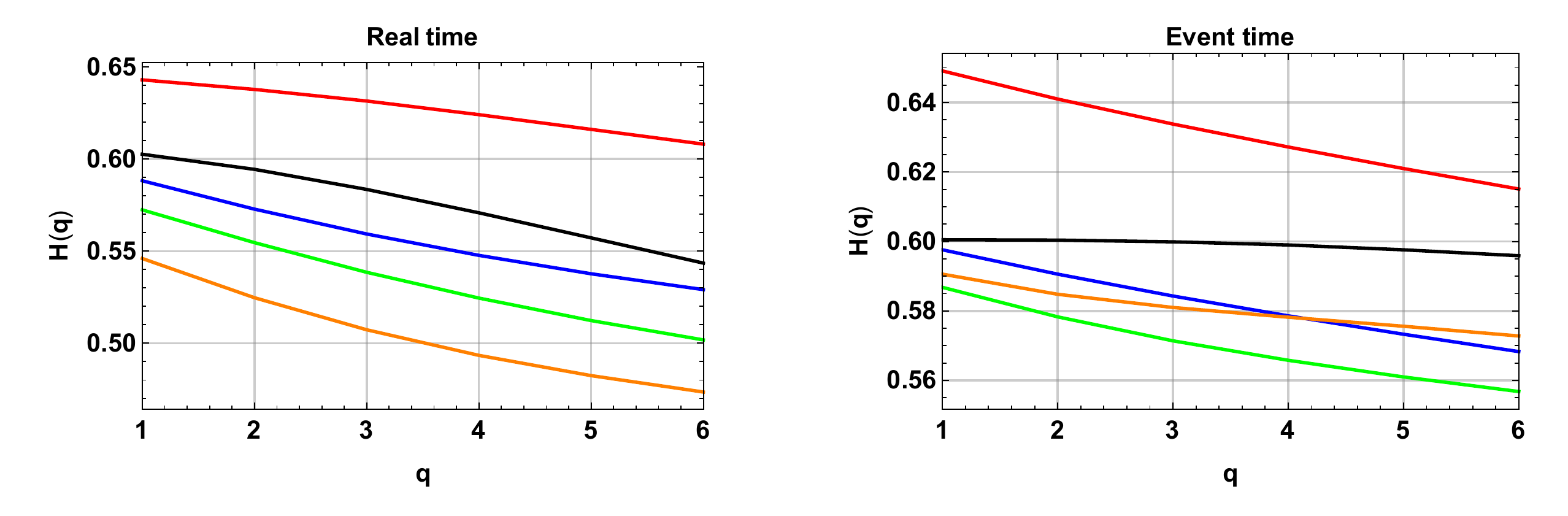}
\par\end{centering}
\caption{Generalized Hurst exponent of order dis-balance real and event time  series.
The generalized Hurst exponent $H(q)$ is calculated using the third method  \eqref{eq:q-fluctuations} for the five stocks: AAPL, AMZN, GOOG, INTC, and MSFT. Here the equal time steps are: $\tau=200 \: s$ for the real-time series and for the event time series $\tau=500 \: ticks$ for the AAPL, AMZN, GOOG stocks, and $\tau=2000 \ ticks$ for the INTC, MSFT stocks. \label{fig4}}
\end{figure}

The multifractal detrended fluctuation analysis is one more essential method providing us the information whether time series are multifractal or mono-fractal. This method applies to the non-stationary time series and can be easily implemented for the analysis of order dis-balance real and event time series. In figure \ref{fig4} we provide results of our calculations described here as the third method.

These results confirm that the order dis-balance real and event time series, at least for the stocks considered, are mono-fractal as functions $H(q)$ are slowly varying and almost linear. For both time series $H(2)$ values are in the interval $0.52 \leq H(2) \leq 0.64$, see table \ref{table2}, and are lower than the values we get by the R/S method. Though the Hurst exponent values for the order dis-balance real and event time series of the investigated stocks are scattered, we have to conclude that R/S and MDFA methods give comparable results and confirm the presence of the long-range memory in the order dis-balance time series. 

\begin{figure}
\begin{centering}
\includegraphics[width=0.95\textwidth]{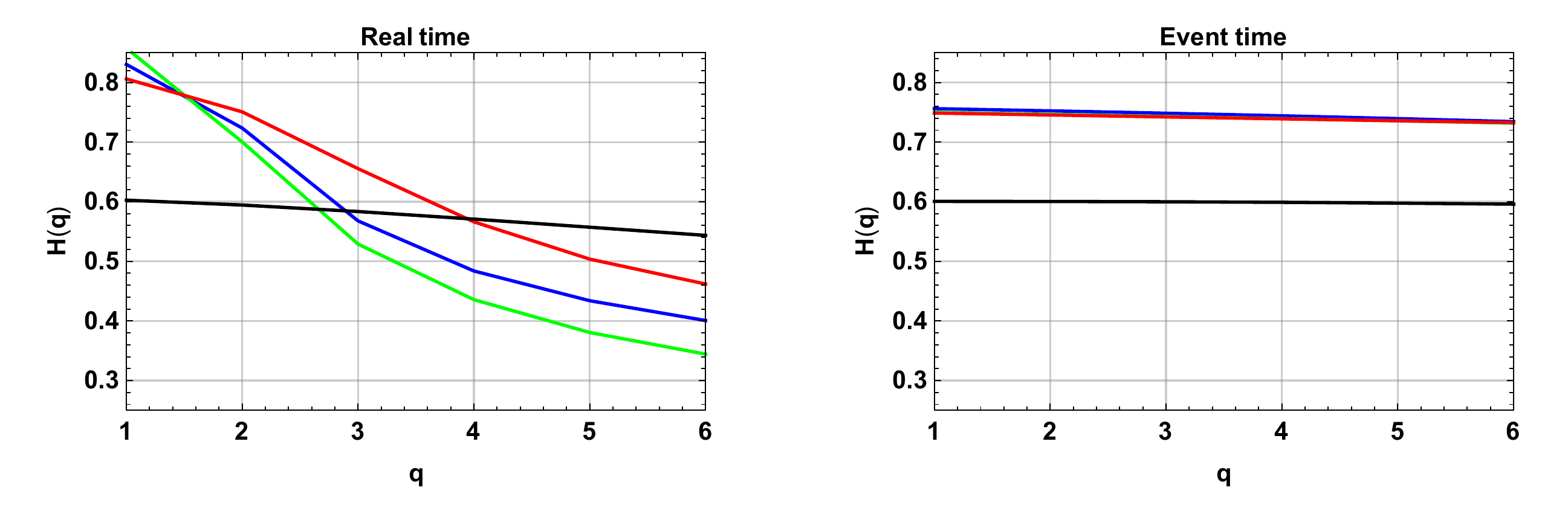}
\par\end{centering}
\caption{The comparison of generalized Hurst exponent of absolute return and order dis-balance time  series.
The generalized Hurst exponent $H(q)$ is calculated by the third method  \eqref{eq:q-fluctuations} for the absolute return and order dis-balance real and event time series of stock AAPL. Here we used the three different time step values for the absolute return and one for the order dis-balance time series. In the real-time sub-figure: (blue) absolute return $\tau=10 \: s$; (green) absolute return $\tau=60 \: s$; (red) absolute return $\tau=200 \: s$; (black) order dis-balance $\tau=200 \: s$. In the event time sub-figure: (blue) absolute return $\tau=200 \: ticks$ ; (green) absolute return $\tau=500 \: ticks$ ; (red) absolute return $\tau=2000 \: ticks$ ; (black) order dis-balance $\tau=2000 \: ticks$. \label{fig5}}
\end{figure}

Many research papers have been devoted to the analyses of multifractality in the financial
data, including absolute return time series \cite{Kantelhardt2002PhysA,Matteo2012PhysA,Green2014EurPhyJB}. The question which properties of the financial time series contribute to the multifractal behavior the most is still open. We compare the generalized Hurst exponent $H(q)$ evaluated for the absolute return and orders dis-balance real and event time series of the stock AAPL in figure \ref{fig5}. The multifractal behavior is evident in the absolute return real-time series and is very weak for the order dis-balance real time series. The event time  series exhibit almost constant $H(q)$ for both absolute return and order dis-balance empirical LOB data for the stock AAPL. 

\begin{figure}
\begin{centering}
\includegraphics[width=0.95\textwidth]{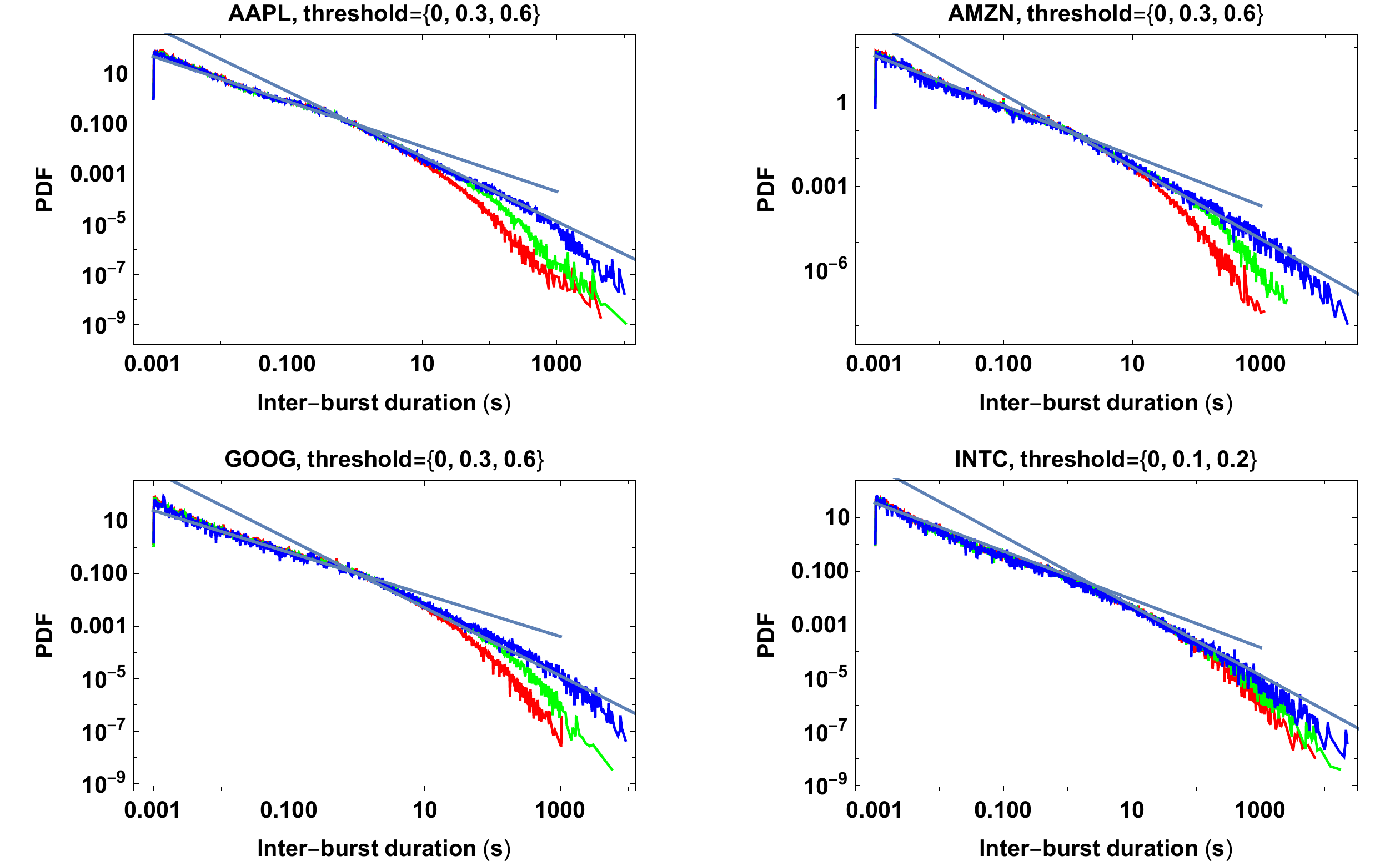}
\par\end{centering}
\caption{Inter-burst duration PDF for the order dis-balance real-time series.
Sub-figures show PDFs of the following stocks: AAPL, AMZN, GOOG, INTC. Values of thresholds are given in the plot labels of sub-figures. The plots are red for the lowest thresholds, green for the higher and blue for the highest. Straight lines, power-laws with exponents $\gamma_1=0.9$ and $\gamma_2=1.3$ guide the eye. \label{fig6}}
\end{figure}

Finally, we present our results of the burst and inter-burst duration analysis. From the R/S and MDFA results, we expect that $H \simeq 0.7$ and the PDF of $T$ should have, at least in some region, a power-law part with the exponent $\gamma_2=2-H \simeq 1.3$. We will denote the exponent of another power-law PDF part by $\gamma_1$ plotting both lines in the PDF of $T$ log-log plot. Thus in both following figures the line $\gamma_2=1.3$ is shown to guide the eye after the power-law we are looking for. In figure \ref{fig6} we demonstrate inter-burst duration, $T$, PDFs for the three different threshold values of order dis-balance real time series of four stocks: AAPL, AMZN, GOOG, INTC.

In the middle part of empirical PDFs, calculated as histograms of $T$ from the time series having up to 25 mln. points, can be fitted very  well by the power-law $T^{-1.3}$. In the region of lower $T$ values PDFs can be approximated by the power-law with exponent $\gamma_1=0.9$ and in the region of the highest values one can observe  exponential like cut-of as is expected for the fBm or Markov processes \cite{Ding1995PhysRevE,Metzler2014Springer,Gontis2020PhysA}. These results confirm that order dis-balance real time series, at least for the stocks considered, have long-range memory characterized by the Hurst exponent approximately equal to $0.7$.

The low value of exponent $\gamma_1=0.9$ probably is related to the additional fluctuations of the order flow intensity. From our point of view, these fluctuations have to be responsible for the PSD in  the region of higher frequencies. Fortunately, there is a method to exclude these fluctuations in the analysis.  

\begin{figure}
\begin{centering}
\includegraphics[width=0.95\textwidth]{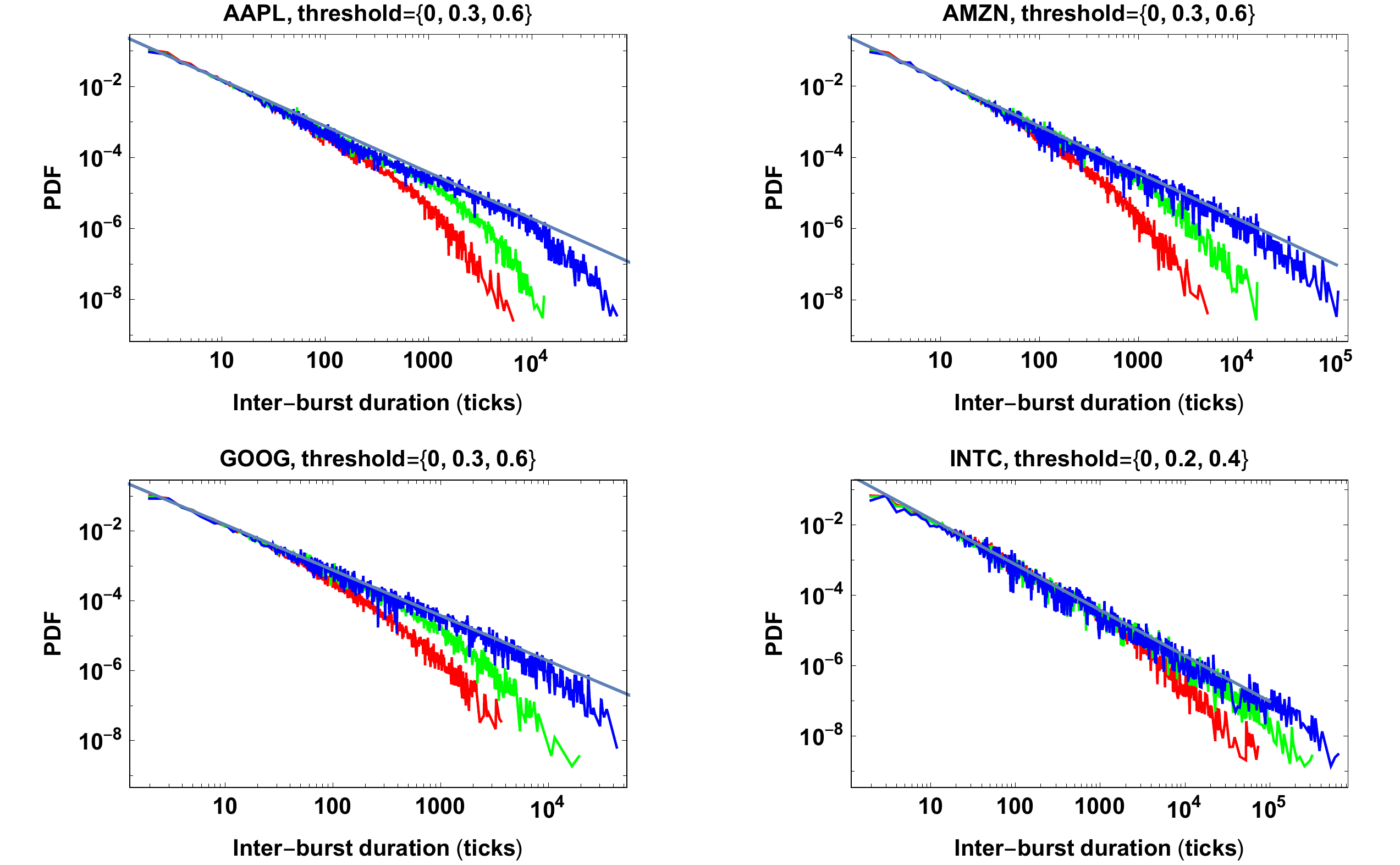}
\par\end{centering}
\caption{Inter-burst duration PDF for the order dis-balance event time series.
Sub-figures show PDFs of the following stocks: AAPL, AMZN, GOOG, INTC. Values of thresholds are given in the plot labels of sub-figures. The plots are red for the lowest thresholds, green for the higher values and blue for the highest. Straight lines, power-laws with exponents $\gamma_2=1.3$, guides the eye. \label{fig7}}
\end{figure}

Seeking to find whether observed long-range memory is related to the properties of order flow intensity, we investigate event time series. In figure \ref{fig7} we demonstrate inter-burst duration, $T$, PDFs for the three different threshold values of order dis-balance event time series for the same stocks as in figure \ref{fig6}. In this case only one power-law with the same exponent $\gamma_2=1.3$ is present. 

This result confirms long-range memory for the order dis-balance event time series with approximately the same Hurst exponent equal to $0.7$, and is even stronger than in the case of real-time series as power-law of PDF spans in the interval of $T$ values up to the five orders. At the first glance, the defined $H$ is very stable, being the same for the all considered 5 stocks, including MSFT and both types of time series. To quantify this stability, we make the numerical least-square log-log fit of $T$ PDF in the observed region of power-law. Numerical results are given in table \ref{table3}.
\begin{table}[!ht]
\centering
\caption{
Hurst exponent $H$ evaluated by the burst and inter-burst duration method for the five stocks order dis-balance real and event time series. Values in parenthesis are calculated from the time series of 2020.}
\begin{tabular}{|l+l|l|l|l|l|l|l|}
\hline
\multicolumn{1}{|l|}{\bf Exponents} & \multicolumn{1}{|l|}{\bf AAPL} & \multicolumn{1}{|l|}{\bf AMZN} & \multicolumn{1}{|l|}{\bf GOOG} & \multicolumn{1}{|l|}{\bf INTC} & \multicolumn{1}{|l|}{\bf MSFT}\\ \thickhline
$H$, real & $0.72(0.69)$ & $0.74(0.73)$ & $0.75(0.62)$ & $0.70(0.70)$ & $0.70(0.69)$ \\ \hline
$\gamma_2$ & $1.28(1.31)$ & $1.26(1.27)$ & $1.25(1.38)$ & $1.30(1.30)$ & $1.30(1.31)$\\ \hline
$H$, event & $0.73(0.71)$ & $0.72(0.72)$ & $0.71(0.60)$ & $0.68(0.67)$ & $0.68(0.68)$ \\ \hline
$\gamma_2$ & $1.27(1.29)$ & $1.28(1.28)$ & $1.29(1.4)$ & $1.32(1.33)$ & $1.32(1.32)$\\ \hline
\end{tabular}
\label{table3}
\end{table}
Numerical values of $H$ are scattered around the values $0.7$ in the narrow region compared to the other methods of $H$ evaluation, see table \ref{table2}. The average value of $H$ for all stocks calculated from the real-time series is $0.72$ and calculated from the event time series $0.7$. These findings serve in a favor for the proposed method to evaluate Hurst exponent from the burst and inter-burst duration analysis. From our point of view, this advantage of the method is related to the dependence of the evaluated $H$ only on the correlations of the noise increments and low sensitivity to the other non-linear origins of power-law statistical properties.
 




\section{Discussion and conclusion \label{sec:conclusion}}

Power-law statistical properties are the characteristic feature of social systems. The financial markets providing us with a vast amount of empirical LOB data exhibit such power-law statistical properties as well \cite{Gould2013QF}. Here we investigated burst and inter-burst duration statistical properties in the order dis-balance time series seeking to develop the test for the presence of true long-range memory. Our previous analysis of burst and inter-burst time statistical properties in the time series of absolute return and trading activity has shown that the observed property of long-range property might be spurious \cite{Gontis2017PhysA,Gontis2018PhysA}. Thus the search for true long-range memory property with correlated stochastic increments in the limit order flow has become of greater interest.

Here we demonstrate, that order dis-balance time series have many peculiarities related to the stock considered. The order flow intensity, the range of order dis-balance fluctuations, the exponents of PSD,  and other measures of long-range memory are varying for different stocks. PSD with two power-law exponents reveals the complexity of these time series. Usually, researchers consider these peculiarities as a result of strong nonstationarity of the financial time series. We have performed the Unit Root (augmented Dickey–Fuller \cite{Dickey1979JASA}) Test, for the tick sizes in the order flow of the considered time series. The unit-roots calculated are less than $10^{-30}$ for all daily intervals of the used empirical data. We cannot detect the change in the regimes of time series as PSD calculated in shorter two-day periods is nearly the same. Even the time series of the same stock in 2012 and 2020 generates very similar PSDs. Differences from stock to stock are more significant than from period to period.

We implemented the second, widely used rescaled range method to evaluate $H$ for the real and event time series. Though the defined values of $H$ are scattered, they are comparable for the real and event time series and with the estimate from the PSD. Fluctuations of Hurst exponent around $H=0.7$ are comparable with defined in order arrivals studied by Lillo \textit{et al} \cite{Lillo2004SNDE,Mike2008JEDC} showing statistically significant variations of the estimated values of
$H$ for different stocks. Despite considerable variations of Hurst exponent the R/S method confirms the long-range memory property. 

The third MF-DFA method applicable to the non-stationary time series gives us additional information on whether the order dis-balance time series are mono-fractal or multifractal. This information is essential for comparing statistical properties observed in the order dis-balance  and absolute return time series \cite{Kononovicius2012PhysA,Matteo2012PhysA}. The MF-DFA implemented for the LOB data confirms that order dis-balance real and event time series are mono-fractal for the considered stocks. The observed dependence of $H(q)$ on $q$ is low and can be compered with the $H(q)$ variations from stock to stock. For the absolute return real-time series extracted from the AAPL LOB data, figure \ref{fig5}, $H(q)$ exhibits much stronger non-linear dependence. The detailed analysis of multifractal spectra is itself effort consuming task \cite{Green2014EurPhyJB}. Here we made very preliminary  multifractal testing of order dis-balance time series concluding mono-fractal behavior. Even for the absolute return, multifractal behavior disappeared when we switched to the event time series, see figure \ref{fig5}. Values $H(2)$ defined by MF-DFA are slightly lower than $H$ defined by the R/S method and confirm that widely used long-range memory estimators give varying results.
\begin{figure}
\begin{centering}
\includegraphics[width=0.95\textwidth]{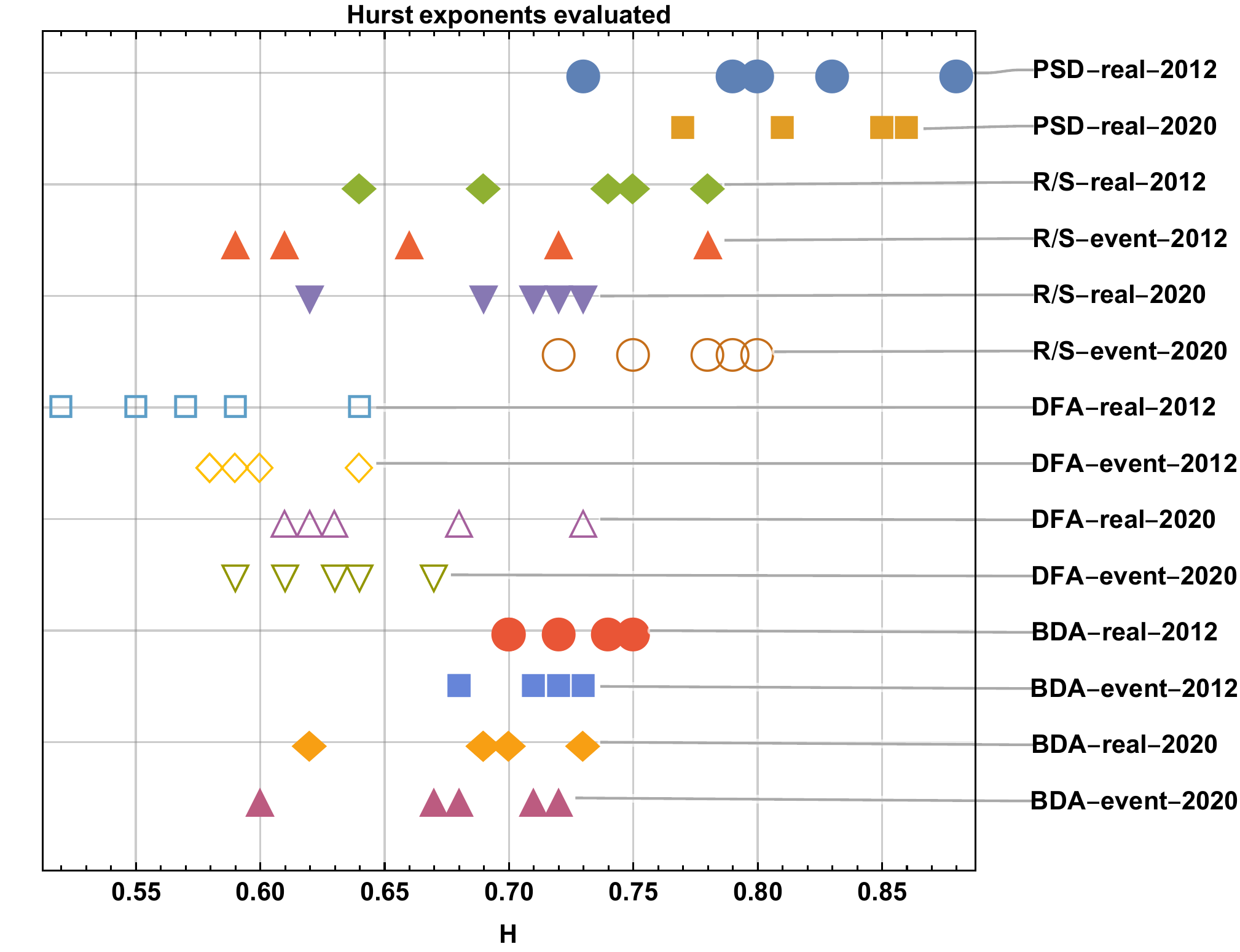}
\par\end{centering}
\caption{Comparison of all used methods to define the Hurst exponent. All rows have $5$ values corresponding to the stocks investigated and are labeled as follows: estimator used (PSD, R/S, DFA, BDA) - time definition (real, event) - the year of series (2012, 2020).  \label{fig8}}
\end{figure}
We investigated statistical properties of burst and inter-burst duration in order dis-balance real and event time series seeking to develop one more test of long-range memory. The big choice of LOB data available and high expectation that the order flow has a real long-range memory property attracted our attention to this type of social system. The results of this empirical analysis are given in figures \ref{fig6}, \ref{fig7}, and table \ref{table3}. All calculated histograms of inter-burst duration for the five stocks investigated can be well-fitted  in the interval of three orders for the real-time series and in the interval of four orders for the event time series by the same power-law with exponent $\gamma=1.3$. Though the more precise numerical evaluation of the exponents $\gamma$, see results in table \ref{table3}, are slightly dispersed, corresponding values of Hurst exponent are in a very narrow region around $0.7$. From our point of view, these results confirm that the limit order flow exhibits true long-range memory, and with high probability, this property may be independent of the stock and time definition. 

Finally, for the comparison, we demonstrate all empirically defined values of Hurst exponent in figure \ref{fig8}, where values are dispersed in the interval $0.5<H<0.9$. Values defined from the burst duration analysis are in the narrower interval $0.67<H<0.75$, except GOOG stock in the 2020 series. PSD and DFA give the most significant deviations from the expected mean of $H=0.7$. It is worth mentioning that real and event time series give similar estimations of the Hurst exponent for all used methods.

Our expectation that burst and inter-burst duration analysis can give an additional value in the estimation of long-range memory property is based on the previous study of the first passage time problem in birth-death processes and the duration PDF in-variance regarding non-linear transformations of the time series \cite{Gontis2012ACS,Gontis2017PhysA,Gontis2017Entropy,Kononovicius2019JStatMech,Gontis2020PhysA}.
This empirical analysis of LOB data strengthens our expectation providing evidence that the definition of Hurst exponent using burst and inter-burst duration analysis can be more reliable in comparison with other widely used methods. From our point of view, it is even more critical that burst duration analysis can give us a method to discern spurious and true long-range memories. Much more extensive study of LOB data in various markets and trading periods is needed to evaluate whether the observed long-range memory is the universal property of the order flow in financial markets. Here let us consider the observed deviation of Hurst exponent defined from BDA in the period 2020 for GOOG as the exception that proves the rule.

\bibliographystyle{iopart-num}
\bibliography{jsm_order-disbalance}

\providecommand{\newblock}{}
\begin{thebibliography}{10}
\expandafter\ifx\csname url\endcsname\relax
  \def\url#1{{\tt #1}}\fi
\expandafter\ifx\csname urlprefix\endcsname\relax\def\urlprefix{URL }\fi
\providecommand{\eprint}[2][]{\url{#2}}

\bibitem{Baillie1996JE}
Baillie R~T 1996 {\em Journal of Econometrics\/} {\bf 73} 5--59

\bibitem{Engle2001QF}
Engle R and Patton A 2001 {\em Quantitative Finance\/} {\bf 1} 237--245
  (\textit{Preprint} \eprint{http://dx.doi.org/10.1088/1469-7688/1/2/305})
  \urlprefix\url{http://dx.doi.org/10.1088/1469-7688/1/2/305}

\bibitem{Plerou2001QF}
Plerou V, Gopikrishnan P, Gabaix X, Amaral L and Stanley H 2001 {\em
  Quantitative Finance\/} {\bf 1} 262--269 (\textit{Preprint}
  \eprint{http://dx.doi.org/10.1088/1469-7688/1/2/308})
  \urlprefix\url{http://dx.doi.org/10.1088/1469-7688/1/2/308}

\bibitem{Gabaix2003Nature}
Gabaix X, Gopikrishnan P, Plerou V and Stanley H~E 2003 {\em Nature\/} {\bf
  423} 267--270

\bibitem{Ding2003Springer}
 2003 {\em Processes with Long-Range Correlations: Theory and Applications\/}
  ({\em Lecture Notes in Physics\/} vol 621) ed Rangarajan G and Ding M
  (Springer) pp XVIII, 398

\bibitem{Gontis2004PhysA}
Gontis V and Kaulakys B 2004 {\em Physica A\/} {\bf 344} 128--133

\bibitem{McCauley2006PhysA}
Bassler K, G G and McCauley 2006 {\em Physica A\/} {\bf 369} 343--353

\bibitem{McCauley2007PhysA}
McCauley J, Gunaratne G and Bassler K 2007 {\em Physica A\/} {\bf 379} 1--9

\bibitem{Micciche2009PRE}
Micciche S 2009 {\em Physical Review E\/} {\bf 79} 031116

\bibitem{Micciche2013FNL}
Micciche S, Lillo F and Mantegna R 2013 {\em Fluctuation and Noise Letters\/}
  {\bf 12} 1340002

\bibitem{Ruseckas2011PRE}
Ruseckas J and Kaulakys B 2011 {\em Phys.Rev.E\/}  051125

\bibitem{Lo1991Econometrica}
Lo A~W 1991 {\em Econometrica\/} {\bf 59} 1279--1313

\bibitem{Willinger1999FinStoch}
Willinger W, Taqqu M~S and Teverovsky V 1999 {\em Finance Stochast.\/} {\bf 3}
  1--13

\bibitem{Mikosch2003}
Mikosch T and Starica C 2003 {\em Long-range dependence effects and ARCH
  modeling, in Theory and applications of long-range dependence\/} (Birkhauser
  Boston)

\bibitem{Ding1993JEmpFin}
Ding Z, Granger C~W~J and Engle R~F 1993 {\em Journal of Empirical Finance\/}
  {\bf 1} 83--106

\bibitem{Bollerslev1996Econometrics}
Bollerslev T and H-O~Mikkelsen H~O 1996 {\em Journal of Econometrics\/} {\bf
  73} 151--184

\bibitem{Giraitis2009}
Giraitis L, Leipus R and Surgailis D 2009 {ARCH}($\infty$) models and long
  memory {\em Handbook of Financial Time Series\/} ed Anderson T~G, Davis R~A,
  Kreis J and Mikosh T (Berlin: Springer Verlag) pp 71--84

\bibitem{Conrad2010}
Conrad C 2010 {\em Journal of Econometrics\/} {\bf 157} 441--457

\bibitem{Arouri2012}
Arouri M~E~H, Hammoudeh S, Lahiani A and Nguyen D~K 2012 {\em The Quarterly
  Review of Economics and Finance\/} {\bf 52} 207--218

\bibitem{Tayefi2012}
Tayefi M and Ramanathan T~V 2012 {\em Austrian Journal of Statistics\/} {\bf
  41} 175--196

\bibitem{Kononovicius2013EPL}
Kononovicius A and Gontis V 2013 {\em EPL\/} {\bf 101} 28001

\bibitem{Gontis2014PlosOne}
Gontis V and Kononovicius A 2014 {\em PLoS ONE\/} {\bf 9} e102201

\bibitem{Gontis2016PhysA}
Gontis V, Havlin S, Kononvicius A, Podobnik B and Stanley E 2016 {\em Physica
  A\/} {\bf 462} 1091--1102

\bibitem{Gontis2017PhysA}
Gontis V and Kononovicius A 2017 {\em Physica A\/} {\bf 483} 266--272

\bibitem{Gontis2018PhysA}
Gontis V and Kononovicius A 2018 {\em Physica A\/} {\bf 505} 1075--1083

\bibitem{Gontis2017Entropy}
Gontis V and Kononovicius A 2017 {\em Entropy\/} {\bf 19} 387

\bibitem{Ding1995PhysRevE}
Ding M and Yang W 1995 {\em Physical Review E\/} {\bf 52} 207--213

\bibitem{Metzler2014Springer}
 2014 {\em First-Passage Phenomena and Their Applications\/} ed Metzler R,
  Oshanin G and Redner S (Springer) p 608

\bibitem{Lillo2004SNDE}
Lillo F and Farmer J 2001 {\em Studies in Nonlinear Dynamics \& Econometrics\/}
  {\bf 8} 1--35

\bibitem{Bouchaud2004QF}
Bouchaud J~P, Gefen Y, Potters M and Wyart M 2004 {\em Quantitative Finance\/}
  {\bf 4} 176--190

\bibitem{Toth2015JEDC}
Toth B, Palit I, Lillo F and Farmer J 2015 {\em Journal of Economic Dynamics \&
  Control\/} {\bf 51} 218--239

\bibitem{Mandelbrot1968SIAMR}
Mandelbrot B and Ness V 1968 {\em SIAM Review\/}  10:422--437

\bibitem{Kaulakys2005PhysRevE}
Kaulakys B, Gontis V and Alaburda M 2005 {\em Physical Review E\/} {\bf 71}
  1--11

\bibitem{Lanouar2011IJBSS}
Lanouar C 2011 {\em International Journal of Business and Social Science\/}
  {\bf 2} 52–66

\bibitem{Beran1994Chapman}
Beran J 1994 {\em Statistics for Long-Memory Processes\/} (Capman I\& Hall)

\bibitem{Kantelhardt2002PhysA}
Kantelhardt J, Zschiegner S, Koscielny-Bunde E, Havlin S, Bunde A and Stanley H
  2002 {\em Physica A\/} {\bf 316} 123119

\bibitem{Peng1994PRE}
Peng C~K, Buldyrev S, Havlin S, Simons M, Stanley H and Goldberger A 1994 {\em
  Physical Review E\/} {\bf 49} 1685--1689

\bibitem{Huang2011Lobster}
Huang R and Polak T 2011 Lobster: The limit order book reconstructor,
  discussion paper school of business and economics, humboldt universit¨at zu
  berlin \urlprefix\url{https://lobsterdata.com/LobsterReport.pdf}

\bibitem{Gontis2012ACS}
Gontis V, Kononovicius A and Reimann S 2012 {\em Advances in Complex Systems\/}
  {\bf 15} 1250071

\bibitem{Gontis2006JStatMech}
Gontis V and Kaulakys B 2006 {\em Journal of Statistical Mechanics\/} {\bf
  P10016} 1--11

\bibitem{Gontis2007PhysA}
Gontis V and Kaulakys B 2007 {\em Physica A\/} {\bf 382} 114--120

\bibitem{Kononovicius2012PhysA}
Kononovicius A and Gontis V 2012 {\em Physica A\/} {\bf 391} 1309--1314

\bibitem{Clauset2009SIAMR}
Clauset A, Shalizi C and Newman M 2009 {\em SIAM Review\/} {\bf 51} 661–703

\bibitem{Corral2006Tectonophysics}
Corral A 2008 {\em Tectonophysics\/} {\bf 424} 177–193

\bibitem{Lennartz2008EPL}
Lennartz S, Livina V, Bunde A and Havlin S 2008 {\em EPL\/} {\bf 81} 69001

\bibitem{Mike2008JEDC}
Mike S and Farmer J 2008 {\em Journal of Economic Dynamics \& Control\/} {\bf
  32} 200--234

\bibitem{Matteo2012PhysA}
Barunik J, Aste T, Di~Matteo T and Liu R 2012 {\em Physica A\/} {\bf 391}
  4234--4251

\bibitem{Green2014EurPhyJB}
Green E, Hanan W and Heffernan D 2014 {\em Eur. Phys. J. B\/} {\bf 87} 1--9

\bibitem{Gontis2020PhysA}
Gontis V and Kononovicius A 2020 {\em Physica A\/} {\bf 540} 123119

\bibitem{Gould2013QF}
Gould M, Porter M, Williams S, McDonald M, Fenn D and Howison S 2013 {\em
  Quantitative Finance\/} {\bf 13} 1709–1742

\bibitem{Dickey1979JASA}
Dickey D and Fuller W 1979 {\em Journal of the American Statistical
  Association\/} {\bf 74} 427–431

\bibitem{Kononovicius2019JStatMech}
Kononovicius A and Gontis V 2019 {\em Journal of Statistical Mechanics: Theory
  and Experiment\/} {\bf 2019} 073402
  \urlprefix\url{https://doi.org/10.1088%2F1742-5468%2Fab2709}

\end{thebibliography}

\end{document}